\pdfoutput=1

\documentclass[11pt]{article}

\usepackage{acl}
\usepackage{amssymb}
\usepackage{times}
\usepackage{latexsym}
\usepackage{graphicx}
\usepackage{longtable}
\usepackage{svg}

\usepackage[T1]{fontenc}

\usepackage[utf8]{inputenc}

\usepackage{microtype}
\usepackage{booktabs}

\title{DrugWatch: A Comprehensive Multi-Source Data Visualisation Platform for Drug Safety Information}

\author{
    Artem Bobrov\textsuperscript{\rm1}$^*$,
    Domantas Saltenis\textsuperscript{\rm2}$^*$,
    Zhaoyue Sun\textsuperscript{\rm2}$^*$, 
    Gabriele Pergola\textsuperscript{\rm2},
\and
    Yulan He\textsuperscript{\rm1,2,3}\\
  \textsuperscript{1}Department of Informatics, King's College London\\
  \textsuperscript{2}Department of Computer Science, University of Warwick \\
  \textsuperscript{3}The Alan Turing Institute\\
  \texttt{\{Artem.Bobrov, Yulan.He\}@kcl.ac.uk} \\
  \texttt{\{Domantas.Saltenis, Zhaoyue.Sun, Gabriele.Pergola.1\}@warwick.ac.uk} \\
  }

\begin{document}
\maketitle
\def\thefootnote{*}\footnotetext{Equal contribution.}
\begin{abstract}

Drug safety research is crucial for maintaining public health, often requiring comprehensive data support. However, the resources currently available to the public are limited and fail to provide a comprehensive understanding of the relationship between drugs and their side effects. This paper introduces \textbf{DrugWatch}, an easy-to-use and interactive multi-source information visualisation platform for drug safety study. It allows users to understand common side effects of drugs and their statistical information, flexibly retrieve relevant medical reports, or annotate their own medical texts with our automated annotation tool. Supported by NLP technology and enriched with interactive visual components, we are committed to providing researchers and practitioners with a one-stop information analysis, retrieval, and annotation service. The demonstration video is available at \url{https://www.youtube.com/watch?v=RTqDgxzETjw}. We also deployed an online demonstration system at \url{https://drugwatch.net/}.


\end{abstract}

\def\thefootnote{\arabic{footnote}}
\section{Introduction}

The use of medications is a cornerstone of modern disease management, yet their potential for adverse reactions can pose safety risks. Adverse drug reactions (ADRs) have been reported to be the most common cause of hospitalisation and rank as the fourth or sixth leading cause of death \cite{lazarou:98}. In addition to the inherent risks of medications themselves, certain drugs may exhibit unpredictable sensitivities in specific patient populations \cite{WHO:04}. Furthermore, there is also a risk of interactions when multiple medications are used concurrently. Therefore, to ensure public health safety, professionals such as physicians, drug developers, and regulatory officials often need to comprehensively understand, assess, and monitor medication safety information from various sources.


To benefit the monitoring of adverse drug events, the World Health Organization (WHO) and numerous countries or regions have established databases for spontaneous case reporting, such as VigiBase \citep{vigibase}, FDA Adverse Event Reporting System (FAERS; \citealt{faers}), and EudraVigilance \citep{eudraVigi}. These databases typically offer interactive query tools to assist users in visualising statistical data from the reporting system. For example, the FAERS Dashboard enable users to search for specific drug products or reaction terms, and offers visual charts illustrating the distribution of corresponding reports by year, demographic details, reaction categories, etc. These databases serve as crucial sources of information for drug safety research.

However, the presence of reports in spontaneous reporting systems does not imply a causal relationship between the drug and the reported adverse reactions. The context in which adverse reactions occur is often complex and may be related to the underlying disease, concurrent medication use, or other factors. Therefore, relying solely on statistical information from spontaneous reporting systems is insufficient for a deeper understanding of drug-induced adverse reactions. Researchers often need access to more detailed information for analysis, much of which is embedded within textual descriptions.

In this paper, we introduce \textbf{DrugWatch}, a multi-source data visualisation platform that integrates information from structured, textual and user-held data on drug safety. It comprises two primary sub-platforms: \textbf{DrugWatch Search} and \textbf{DrugWatch Annotate}. \emph{DrugWatch Search} offers users visualised statistical data sourced from FAERS and PubMed, along with robust support for fine-grained PubMed medical case report retrieval. \emph{DrugWatch Annotate} empowers users to annotate their private data and visualise the resulting annotations.

For \emph{DrugWatch Search}, similar to the FAERS Dashboard, we enable users to search for drugs or adverse reactions and visualise the statistical information provided by the FAERS Database. However, we additionally utilise event extraction techniques to retrieve textual context and present statistics extracted from text data for user queries. We gather medical case reports related to ADRs from PubMed and extract structured information about adverse events using the approach proposed by \citet{sun2024leveraging}. We present the statistical information of the extracted results in a similar way to that of the FAERS data for easy comparison. Additionally, we provide users with a list of PubMed literature and abstracts associated with their search queries, enabling them to conveniently access detailed descriptions of the events in medical texts. Users can also customise more granular search criteria, such as limiting results based on patient age or gender, to filter the search results.

For \emph{DrugWatch Annotate}, we integrate several pre-trained models such as Flan-T5 \citep{chung2022scaling} and UIE \citep{lu-etal-2022-unified}, and an LLM, i.e., Mistral-7B \citep{mistral-7b}, that enables users to perform fine-grained pharmacovigilance event extraction on their private data. These models support the extraction of \emph{subject}, \emph{treatment}, and \emph{effect} information for adverse drug events (ADEs) and potential therapeutic events (PTEs), along with their sub-arguments (e.g., demographic information and drug administration details). We support the visualisation of annotation results, allowing users to quickly try a single data point through a demo window or batch-view the visualised annotations for each data entry. We have pre-processed and visualised manual annotations from the PHEE dataset \citep{sun-etal-2022-phee} and different model predictions for direct model comparison and selection.


The contributions of this paper can be summarised as follows:
\begin{itemize}
    \item We propose a multi-source drug safety information visualisation platform, facilitating users to perform comprehensive analysis on structured data from spontaneous case reports and textual data from medical literature or private sources.
    \item Our platform supports flexible retrieval mechanisms, allowing users to obtain statistics visualisations based on different search items and compare data from different sources. We also integrate a text retrieval system based on event extraction, enabling users to retrieve textual evidence from medical literature.
    \item We allow users to perform pharmacovigilance event extraction and visualise annotation results on their private data, offering a range of models for their selection.
\end{itemize}



\section{Architecture of DrugWatch}

\begin{figure}[ht]
    \centering
    \includegraphics[width=0.46\textwidth]{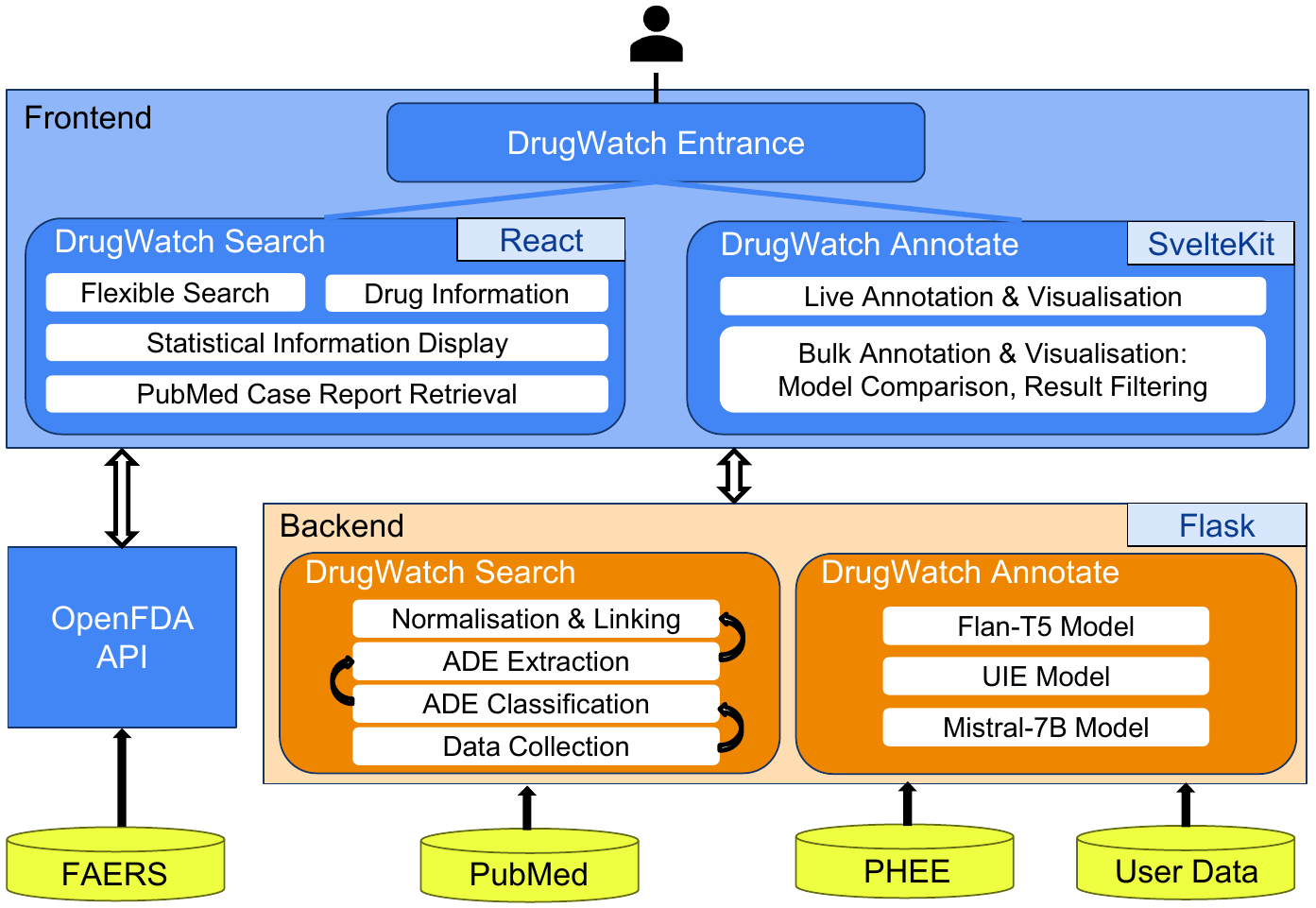}
    \caption{The overall architecture of \textit{DrugWatch}.}
    \label{fig:architecture}
\end{figure}

DrugWatch consists of two sub-platforms: \emph{DrugWatch Search} and \emph{DrugWatch Annotate}. The overall architecture is illustrated in Figure \ref{fig:architecture}. 

\emph{DrugWatch Search} is designed for flexible drug and ADE search. It presents not only fundamental information about drugs but also statistical information from the FAERS database and PubMed literature collections. Additionally, it enables users to access relevant ADE case reports seamlessly. Its front-end is implemented using React \cite{ReactJS}, creating a smooth and interactive user experience. On the server side, we utilise the Flask \cite{Flask} framework to manage API requests and handle data processing from local databases. 

\emph{DrugWatch Annotate} provides automated prediction and visualisation services for user-held data. Users can instantly or in bulk extract ADEs from their data using our built-in fine-tuned models or LLMs. They can easily visualise the event arguments extracted for each data instance, compare prediction results from different models, and conveniently filter results. We preload the PHEE dataset for direct comparison purposes. The front-end of DrugWatch Annotate is built with SvelteKit \cite{svelte}, ensuring fast responsiveness and a clear, visually appealing user experience. The backend continues to utilise Flask. 






\section{User Interaction Design}


\subsection{DrugWatch Search Sub-platform} 

\emph{DrugWatch Search} is a search-centric multi-source information display platform that allows users to search for drugs or side effects flexibly. It not only presents basic drug information for users but also supports the visualisation of interactive statistical information. The integrated event extraction algorithm further enables the platform to retrieve and display relevant PubMed literature based on flexible search options.

\paragraph{Flexible Search} Users can search for individual drugs or side effects and their respective combinations. A combination search for drugs and side effects is currently not available from the search entrance, but users can filter search results by side effects (or drugs). Furthermore, for any queried drug or side effect, the platform provides the option to refine search results based on specific demographic filters, including patient gender, age group (or exact age), and nationality. For a visual guide, see Figure \ref{fig:search_page} and Figure \ref{fig:search_options} in Appendix.

\paragraph{Drug Information Display} When searching for drugs, our platform first presents users with basic drug information collected from DrugBank. This includes structural diagrams, IUPAC name, chemical class, and chemical formula of the drug molecule. Additionally, we display information related to drug use such as indication, half-life, and brand names. When users query multiple drugs at once, the information for each drug is displayed sequentially. See Figure \ref{fig:drug_info} for an illustration.

\paragraph{Statistical Information Display} We provide statistical information for reports meeting customised search criteria on the main results page, and offer a breakdown of demographic information of the searched drug or side effects in a pop-up window. For both cases, users can compare information from FAERS and PubMed with a single click.

\begin{figure}[h]
    \centering
    \includegraphics[width=1\linewidth]{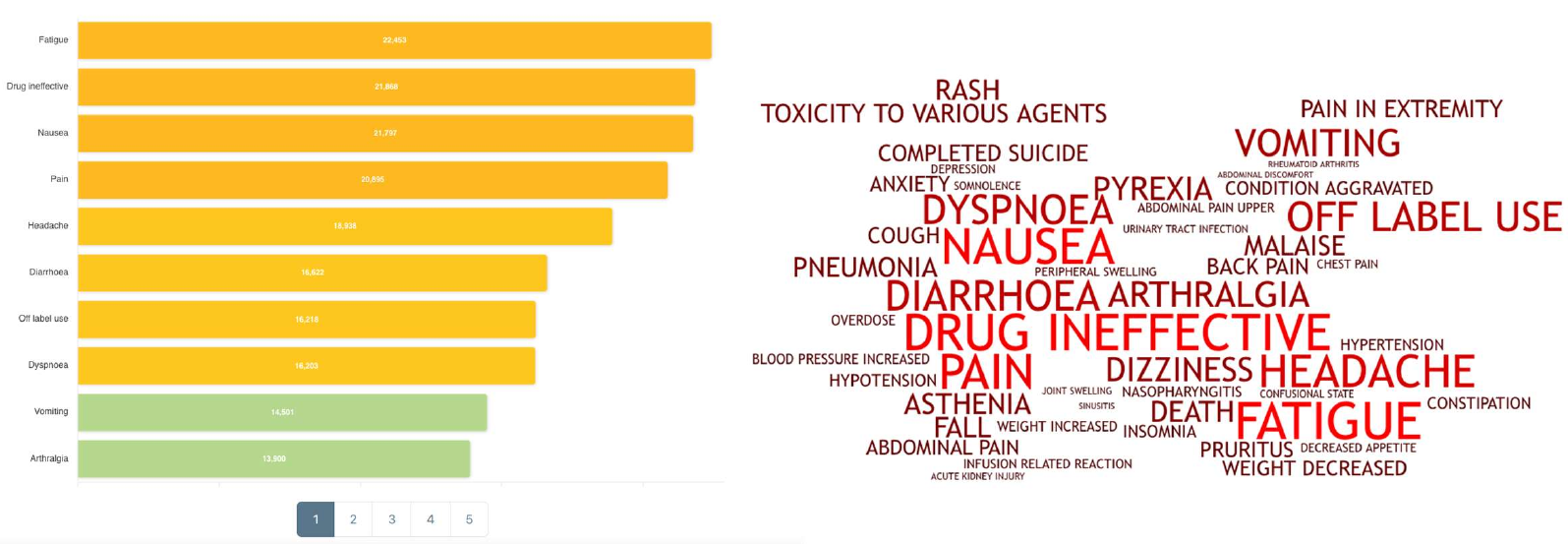}
    \caption{Top frequent side effects related to Acetaminophen, presented by bar chart and word cloud.}
    \label{fig:term_chart_word_cloud}
\end{figure}

On the main results page, we initially display a line chart (Figure \ref{fig:time_series}) showing the variation in the number of reports matching the search criteria over the years. Additionally, we present the most relevant side effects (or drugs) associated with the drug (or side effect) queried by the user. We present this information in two different ways. Firstly, users can observe the top 50 side effects (or drugs) with the highest frequency in the reports, along with their respective counts and proportions (by mouseover), through a bar chart. For easier viewing, the results are divided into 5 pages, with different colours indicating the rarity of the terms. Secondly, users can visually grasp the distribution of related terms through a word cloud. Figure \ref{fig:term_chart_word_cloud} illustrates examples of these two approaches.

\begin{figure}[!h]
    \centering
    \includegraphics[width=0.9\linewidth]{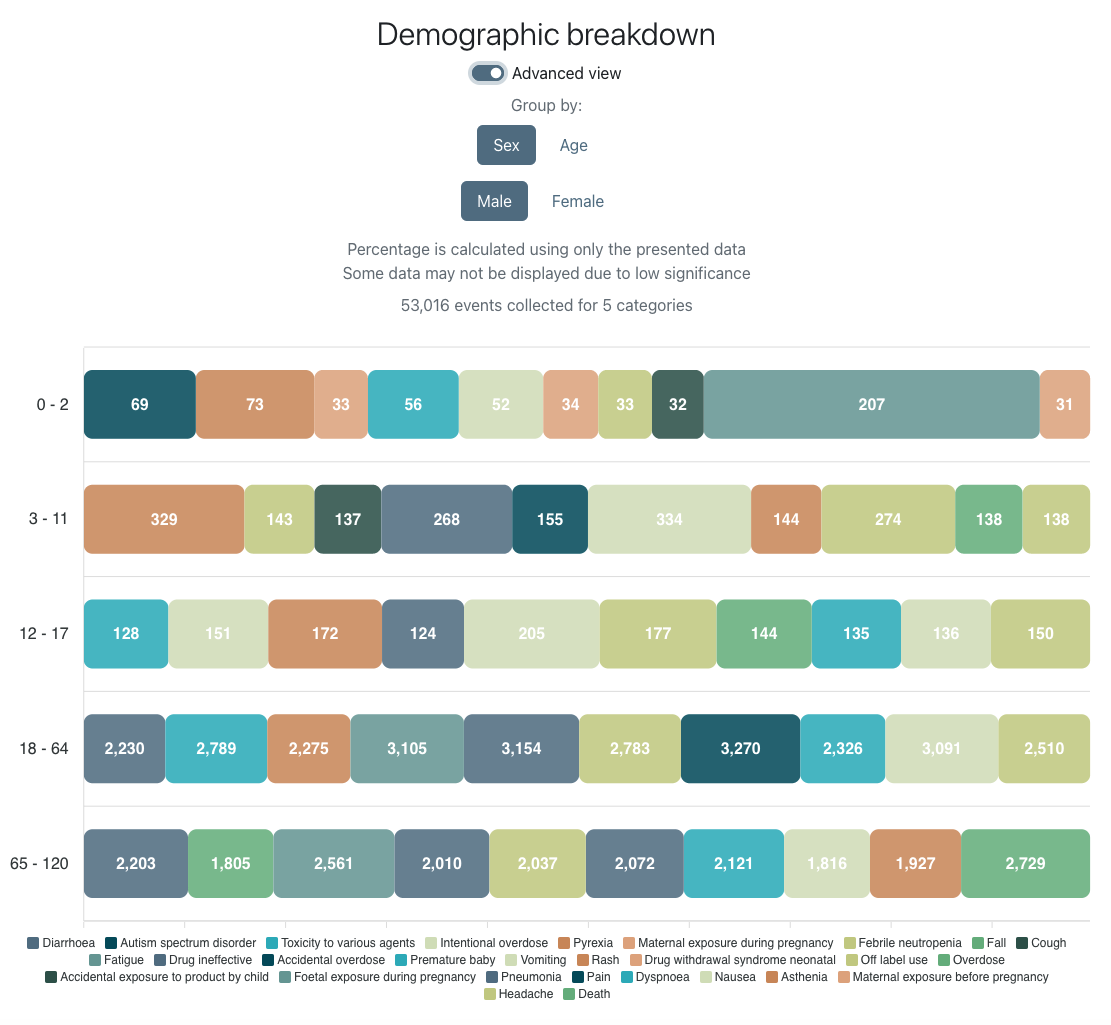}
    \caption{A breakdown of top side effects for each age group when searching for reports of Acetaminophen in males.}
    \label{fig:demo_breakdown}
\end{figure}

The demographic information page first displays the comprehensive age and gender distribution of all reports linked to the queried drug or side effect through a pie chart, facilitating users in visually perceiving the distribution across different demographic groups (Figure \ref{fig:demographic_distribution}). Additionally, we provide a bar chart for any age or gender group, exhibiting the quantities and proportions of the top 10 reported side effects or drugs (Figure \ref{fig:breakdown_basic}). Should users seek further insights into age and gender group comparisons, our advanced view supplies detailed counts of each top side effect or drug-related reports within age groups across gender groups (or in reverse). Figure \ref{fig:demo_breakdown} shows a screenshot of the demographic breakdown charts.

\begin{figure}[!h]
    \centering
    \includegraphics[width=1\linewidth]{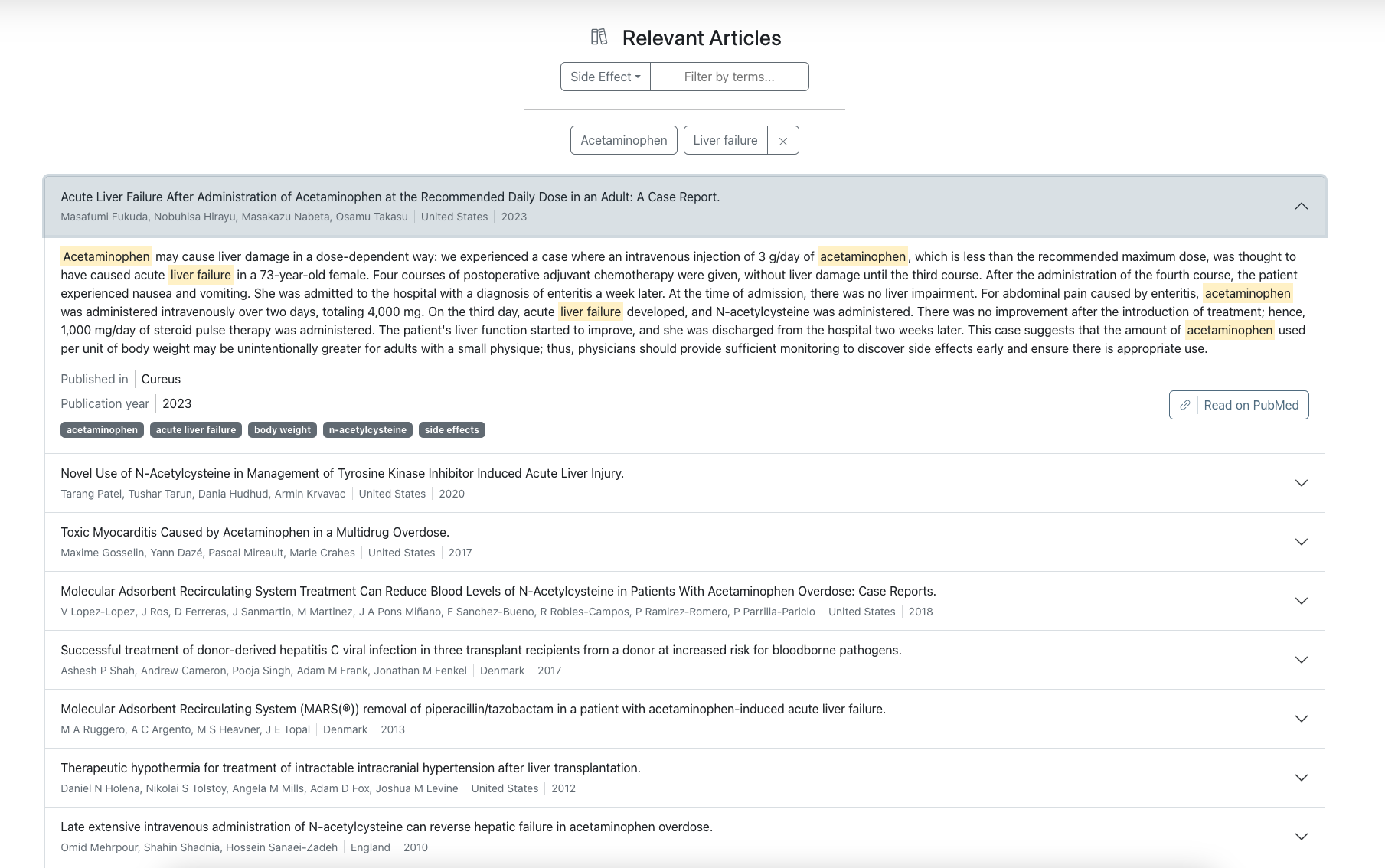}
    \caption{Retrieved PubMed case reports related to "Acetaminophen" and "Liver failure".}
    \label{fig:relevant_articles}
\end{figure}

\paragraph{Pubmed Case Report Retrieval} We simultaneously present users with relevant PubMed case reports on the search results page, allowing them to quickly grasp contextual information surrounding the adverse drug reactions. By default, we display information such as the literature titles, abstracts, keywords etc., that match the search criteria, along with links for quick access to the respective PubMed entries. Users' search terms are highlighted in the abstracts for easier browsing. In addition, we have designed a flexible interaction method, allowing users to dynamically adjust the literature search criteria as needed. For example, they can interact with the bar chart or word cloud chart depicting the distribution of adverse reactions on the search page to obtain a list of literature associated with both the searched drug and an adverse reaction. They can also specify any other filtering terms by manual input (as shown in Figure \ref{fig:relevant_articles}).

\subsection{DrugWatch Annotate Sub-platform}

\emph{DrugWatch Annotate} provides users with automated annotation capabilities for adverse events, facilitating the visualisation of annotated results. It features a live annotation interface for real-time check of individual data entries and a bulk annotation interface for efficient assessment of uploaded batch data. The annotation platform lays the foundation for users to conduct in-depth analysis of their private data subsequently.

\paragraph{Live Annotation} Users can freely input sentences they wish to analyse into the text window and then select the model they want to apply. Currently, three models (i.e., Flan-T5, UIE, Mistral-7B) are available for annotation. The page displays visual results of the model's predictions in real-time, presenting arguments in different colours for easy browsing. Users can also view and copy the results in JSON format (as shown in Figure \ref{fig:live_annotate}). 

\begin{figure}[!h]
    \centering
    \includegraphics[width=1\linewidth]{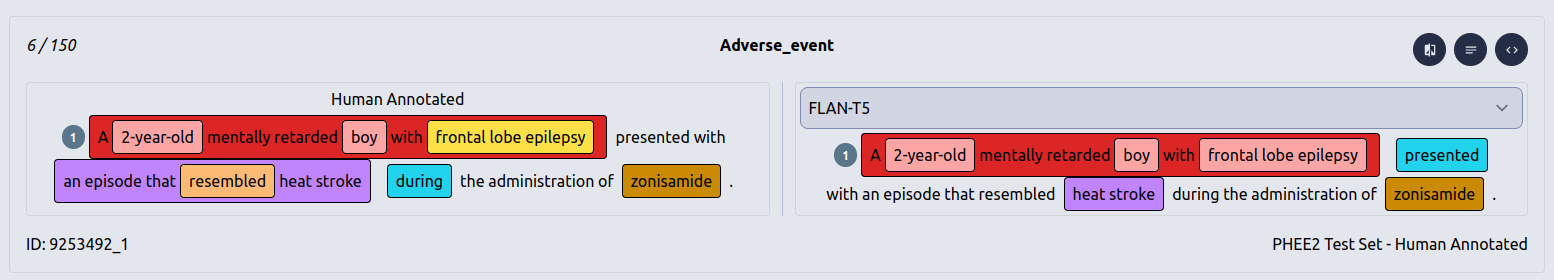}
    \caption{Illustration of \textit{DrugWatch Annotate} model annotation result comparison interface. }
    \label{fig:annotate_model_compare}
\end{figure}

\paragraph{Bulk Annotation} Users can upload their data in batches and visualise all annotation results. They may view annotation results for any single model or compare results from two models side by side (Figure \ref{fig:annotate_model_compare}). We default to loading the PHEE dataset and specifically provide manual annotations on this dataset for users to compare the effects of different models and make selections accordingly. We also allow users to search annotation results to only view results containing a specific argument type (e.g., subject's age) or containing a specific argument span (e.g., "6 years old"). Similarly, users may check and export model outputs in JSON format. A more comprehensive illustration of the bulk annotation interface is shown in Figure \ref{fig:bulk_annotate}.

\section{Backend Implementation}

\subsection{Data Storage and Retrieval} For \emph{DrugWatch Search}, to retrieve statistical data or article information from the back-end, the front-end service makes a series of REST API requests to various endpoints implemented in our backend. Data from the FAERS system is directly fetched from the front-end through the OpenFDA API\footnote{\href{https://open.fda.gov/}{https://open.fda.gov/}}, while text and statistical information from PubMed are processed and stored in the local file system and retrieved from the back-end. This integrated approach effectively reduces the server's workload and ensures the speed. Specifically, the PubMed data are stored in JSON files and loaded into RAM to expedite the search process. We transform the results of the extraction method described in subsection \ref{sec:pubmed_integration} and remove fields that will not be used in the search. Once a search request is submitted, the algorithm iterates through the extracted event arguments of the loaded data, and if a match is found, appends the article ID to the response. Finally, the metadata for each retrieved article is fetched with \textit{Biopython}\footnote{\href{https://biopython.org}{https://biopython.org}} and returned to the front-end.

For \emph{DrugWatch Annotate}, in order to protect data privacy, we do not store user inputs or uploaded documents on the platform. These data are only retained while the user session is active, meaning that reloading the page will clear the uploaded data from the session. After data is uploaded, a request with the provided data is sent to the back-end, which then reads, annotates, and returns the results for visualisation on the front-end. However, this implies that users may need to wait for the model processing. Additionally, we have preprocessed the annotated PHEE dataset, storing both manually annotated data and model prediction results in a local PostgreSQL database\footnote{\href{https://www.postgresql.org/}{https://www.postgresql.org/}}, allowing users to directly access existing datasets and save time.

\subsection{Relevant Medical Case Reports Integration}
\label{sec:pubmed_integration}

To integrate user query-related medical case reports from PubMed into our platform, we first retrieve and download abstracts of adverse event case reports from PubMed, and train a classifier to filter sentences mentioning adverse events from these abstracts. Subsequently, we utilise an event extraction model to extract fine-grained event arguments from the filtered sentences and store the extraction results and their relevant PubMed IDs locally. We then normalise the extracted arguments using regularisation methods, match them with user queries and return relevant PubMed IDs. Finally, we provide a preview of the list and abstracts of the retrieved articles.

\paragraph{Case Reports Collection} The initial stage involves obtaining abstracts from PubMed that pertain to adverse events. We use \textit{Biopython}
to fetch data from PubMed. During retrieval, we obtain records containing the keywords "adverse event", "adverse effect", "adverse reaction", or "side effect", while restricting the publication type to "case report", the language to English, and the presence of an abstract. Our platform has currently collected and analysed case reports up to December 2023, and allows for incremental data updates over time. In total of \textasciitilde{184k} articles are collected at this stage. 

\paragraph{Sentence Classification} To extract more granular adverse event information, we first filter out sentences mentioning adverse events from the collected abstracts. We train a binary classifier based on SciBert \citep{beltagy-etal-2019-scibert} and apply it to all sentences. The classifier was trained on the ADE dataset \citep{gurulingappa2012development}. Around 78k publications, which contain 220k sentences related to ADEs, remained after classification.

\paragraph{Adverse Event Extraction} We then extract fine-grained structured information from the selected sentences related to ADEs, including drug names, adverse reactions, drug administration information, patient demographic information, etc. These extracted arguments are later used to support flexible retrieval functionalities. We utilise the fine-tuned Flan-T5 model introduced in our previous work \citep{sun2024leveraging} to extract arguments of adverse events sentence by sentence. The model converts structured event information into linearised text sequences to train a Seq-to-Seq model on the PHEE \citep{sun-etal-2022-phee} dataset.

\paragraph{Result Normalisation and Linking} The final step is to map the extracted results to the user's query and return the corresponding publication links and abstracts. The results of event extraction are first transformed into structured data, removing fields that are not relevant for search, and merging extraction results from the same article by drug. Subsequently, as the text-based extraction results are free-text spans with rich expressions, we associate them with search terms through normalisation. Specifically, for age and gender fields, we collected all expressions appearing in the database, mapped them to a range or specific value using GPT-4, and verified them manually. For drug and side effect terms, we cleaned them using regular expressions based on manual rules and dictionaries. Finally, when receiving a query, we traverse the entire database to search for matching arguments and return the associated PubMed IDs.


\subsection{Annotation Models}
We provide several models for the user to annotate pharmacovigilance events of their own data through the DrugWatch Annotate platform. For the fine-tuned models, we utilise the UIE and Flan-T5 models trained by \citet{sun2024leveraging} on the PHEE dataset, which have been reported to achieve good performance and are easy to use. The application for these models is similar to that described in subsection \ref{sec:pubmed_integration}. For the LLM, we supply Mistral-7B \citep{mistral-7b}. We deploy the model on our local server and perform inference using the `llama.cpp'\footnote{\href{https://github.com/ggerganov/llama.cpp}{https://github.com/ggerganov/llama.cpp}} library. 
We configure a grammar file to restrict the model's output to JSON format and provide an ADE and a PTE example as demonstrations. However, format corruption still commonly exists in the model's output. To address this issue, we further apply the untruncateJson\footnote{\href{https://github.com/dphilipson/untruncate-json}{https://github.com/dphilipson/untruncate-json}} library to complete the JSON output from the model to be parsable format.



\section{Model Evaluation}

\paragraph{ADE Classification Evaluation}

\noindent The classifier is trained and evaluated on the ADE dataset \citep{gurulingappa2012development}. It contains approximately 4k ADE-related sentences and 16k negative sentences. We sample an equal number of negative sentences as the positive ones for training and evaluation. 
The data is split by 7/1/2 for training/validation/testing. The classification evaluation result is presented in Table \ref{tab:cls-result}. 

\begin{table}[h]
\small
\centering
\resizebox{0.97\columnwidth}{!}{
\begin{tabular}{@{}cccc@{}}
\toprule
   P(\%) & R(\%) & F1(\%) & Accuracy(\%) \\ \midrule
90.56          & 93.21             & 91.86          & 91.74            \\ \bottomrule
\end{tabular}}
\caption{ADE classifier evaluation result.}

\label{tab:cls-result}
\end{table}

\paragraph{ADE Extraction Evaluation}
\noindent We use the PHEE \citep{sun-etal-2022-phee} dataset to train and evaluate the event extraction model. The dataset contains annotations for PTEs and ADEs, and hierarchically annotates the main arguments and sub-arguments of the events. In total around 5k sentences are included in the dataset, and are split by 6/2/2 for training/validation/testing. 

Table \ref{tab:overall-arg-result} presents the performance of our event extraction model applied to DrugWatch Annotate. Here, $EM\_F1$ measures the exact match of the argument span, while $Token\_F1$ measures the matched tokens in the arguments. Constrained by the available hardware and the size of models that can run on it, the performance of the LLM (Mistral-7B) still lags far behind fine-tuned smaller models. After upgrading our equipment in the future, we will deploy more powerful models which may result in better performance.
\begin{table}[h]
\small
\centering
\resizebox{0.97\columnwidth}{!}{
\begin{tabular}{@{}lcccc@{}}
\toprule
  & \multicolumn{2}{c}{Main-arguments} & \multicolumn{2}{c}{Sub-arguments} \\ \cmidrule(l){2-3} \cmidrule(l){4-5} 
   & EM\_F1 & Token\_F1 & EM\_F1 & Token\_F1 \\ \midrule
Flan-T5 (Large)       & \textbf{71.13}          & \textbf{83.40}             & \textbf{77.43}          & \textbf{78.97}            \\
UIE (Large)       & 70.02          & 81.88             & 75.25          & 76.52            \\
Mistral (7B)       & 38.97          & 50.43             & 32.33          & 33.00           \\
\hline
\end{tabular}}
\caption{Overall argument extraction results of integrated models in DrugWatch Annotate. }

\label{tab:overall-arg-result}
\end{table}

\begin{figure}[h]
\centering
\includegraphics[scale=0.25]{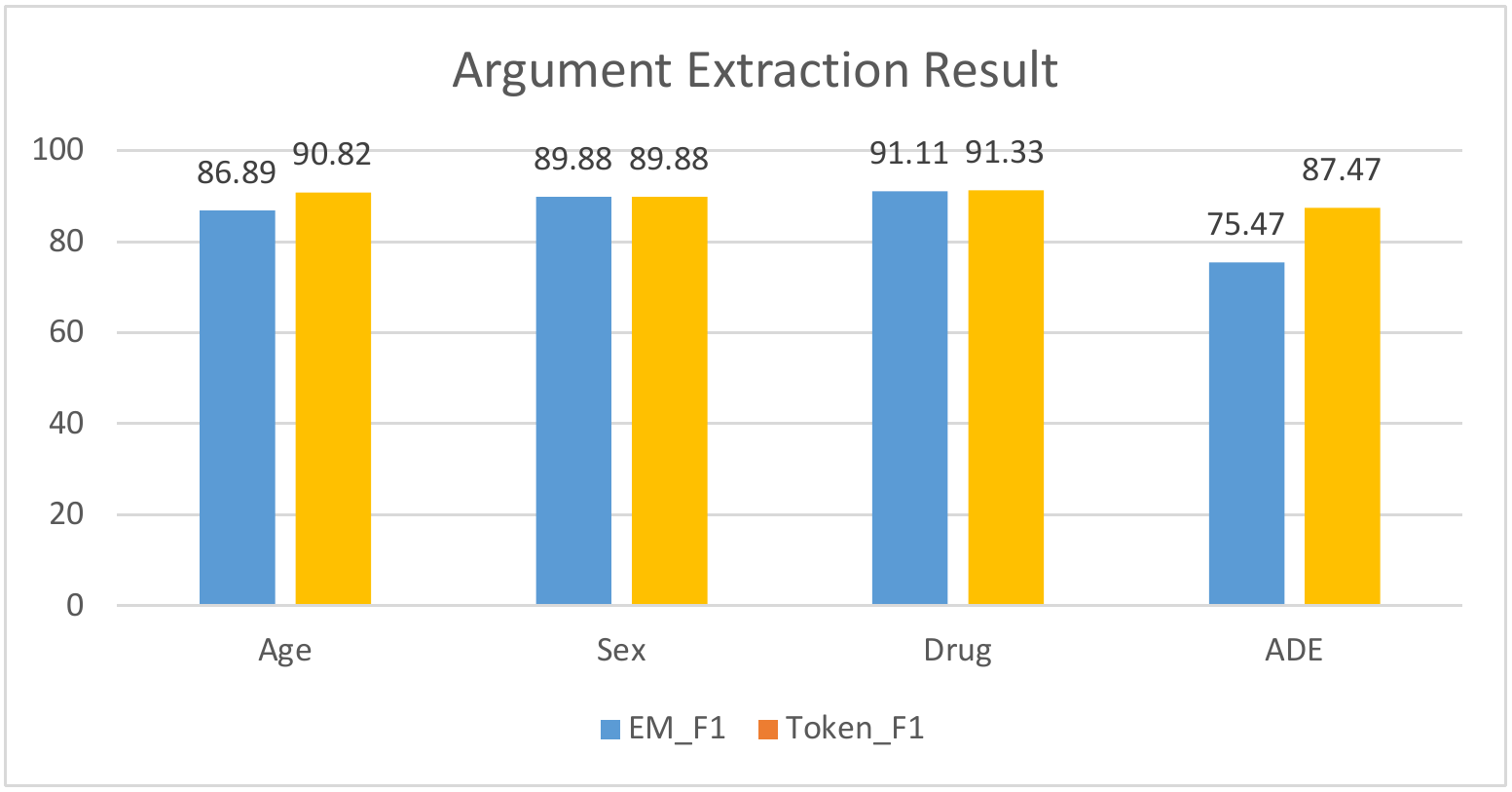} 
\caption{Extraction results of the Flan-T5 model for certain arguments of ADEs used by \emph{DrugWatch Search}.}
\label{fig:arg-rst}
\end{figure}

We employ the Flan-T5 model for event extraction in DrugWatch Search. We keep the model trained on the original data but only use partial results as needed. Specifically, we leverage the model to extract the patient's age, gender, treated drugs, and adverse effects related to adverse events. Figure \ref{fig:arg-rst} shows the corresponding extraction performance for these arguments.


\section{Conclusion and Future Work}

DrugWatch is a multi-source data visualisation and annotation platform for drug safety research. It aims to provide a comprehensive, interactive information retrieval experience. We are committed to alleviating the inconvenience researchers often face when navigating multiple platforms to access information on drug adverse reactions. To achieve this, we integrate statistical and textual information from the spontaneous case report systems (i.e., FAERS) and medical literature databases (i.e., PubMed), allowing users to conduct interactive searches. We also support users to annotate and visualise their own text, laying the foundation for subsequent in-depth private data analysis.

In future work, we will consider further expanding data sources and supporting more granular searches. In particular, we currently do not support statistical analysis of users' private data due to data security considerations, which is an issue we are actively working to address. Furthermore, we are considering integrating literature summaries or question-answer components into the system to support summarisation or questioning of retrieved PubMed texts, enabling users to learn diverse information seamlessly.

\section*{Limitations}
\label{sec:limitations}

DrugWatch Search currently does not support searching for a combination of drugs and adverse reactions, e.g., users cannot search for both "Acetaminophen" and "nausea" simultaneously. This limitation arises from restriction calling OpenFDA API (i.e., specific search types must be specified) and the display logic. To compensate for this, we allow users to conveniently navigate to demographic information pages related to associated side effects when searching for drugs, and vice versa. Additionally, through dynamic article searching, filtering retrieved PubMed literature by combined search terms can be performed.

In addition, our visualisation of FAERS data relies on the OpenFDA API. Therefore, when API access is restricted or reaches its limit, it may fail to display information from the FAERS database.

Furthermore, when users utilise DrugWatch Annotate for batch data prediction, considering the sensitivity of medical data, we avoid storing user data on the server side. This means that all data processing will occur within a single session, and users may need to wait online for processing results. The duration of the wait depends on the server hardware infrastructure and the amount of text uploaded by the user.

Moreover, users currently can only upload data in the format specified by us, which is a text file with one sentence per line. Additionally, the extracted events must adhere to our predefined schema. In the future, with the integration of more powerful LLMs, we will allow users to customize the structure of the events they want to extract.





\section*{Ethics Statement}

Neither spontaneous case reports from FAERS nor medical reports from PubMed suggest a direct causal relationship between the drug and the adverse effect. Users should avoid relying on our platform to make healthcare decisions. Particularly, although we provide visualisations of PubMed case report statistics, users should note the sparse nature of ADE reports from medical literature and the potential for statistical bias therein.

In terms of user data annotation, while we prioritise the security of user data and refrain from storing any on our servers, these data must still transmit through the network, posing inherent risks. Users should acknowledge these risks and consider using de-identified or synthetic data when starting with our platform.

\section*{Acknowledgements}
The authors would like to express their gratitude for the valuable insights and feedback provided by Dr. Bino John, Dr. Nigel Greene, and Dr. Matthew Arnold from AstraZeneca. This work was supported in part by the UK Engineering and Physical Sciences Research Council through a Turing AI Fellowship (EP/V020579/1, EP/V020579/2).

\bibliography{custom}
\bibliographystyle{acl_natbib}

\newpage
\appendix

\setcounter{table}{0}
\renewcommand{\thetable}{A\arabic{table}}
\setcounter{figure}{0}
\renewcommand{\thefigure}{A\arabic{figure}}

\section{Related Work}

Several resources and tools have already been available to researchers and practitioners in pharmacovigilance for understanding adverse drug reactions. One important category is spontaneous reporting systems, which allow patients, physicians, or other practitioners to spontaneously submit ADE case reports to the database. These databases include VigiBase \citep{vigibase}, FAERS \citep{faers}, and EudraVigilance \citep{eudraVigi}, etc. They collect large amounts of structured information on ADEs, serving as primary sources for adverse reaction monitoring. Some spontaneous reporting systems are also equipped with sophisticated visualisation tools, e.g., FAERS Dashboard, helping users visualise statistical information related to adverse reactions in graphical form. However, these charts only provide an overview of adverse reactions from a data perspective, while specific descriptions of adverse reactions including their causes are often hidden in texts. This requires researchers to search for additional literature to learn more detailed information about the adverse event. Therefore, we are committed to integrating text information retrieval with statistical information visualisation on the same platform, providing convenient and unified interactive design to save users time across platforms.

Another useful category of resources is knowledge bases related to drugs and adverse reactions. Among them, DrugBank \citep{DrugBank} provides detailed information on drug pharmacology and properties et al., but adverse reaction data is not publicly available and is only presented in structured data tables for known adverse reactions. The SIDER \citep{SIDER_KUHN_2015} database is open-sourced and provides drug-adverse reaction pairs extracted from drug package inserts. Another platform with a similar intention to ours, also dedicated to comprehensive adverse reaction information services, is MetaADEDB \citep{meta_ade_db}. However, its design resembles more of a knowledge base, presenting known knowledge including synonyms, indications, and ADRs ever reported in the FAERS system. MetaADEDB 2.0 also includes a prediction system, but it focuses on molecular structure-based ADR prediction, which is different from our text-based event extraction tool. The main difference between these knowledge-based tools and our platform is that they focus on existing knowledge, are infrequently updated, and limited support for visualisation; whereas our platform focuses on real, specific adverse reaction events, is regularly updated, and provides rich and interactive visualisation interfaces.

\section{UI Supplementaries}

\begin{figure}[!h]
    \centering
    \includegraphics[width=0.3\textwidth]{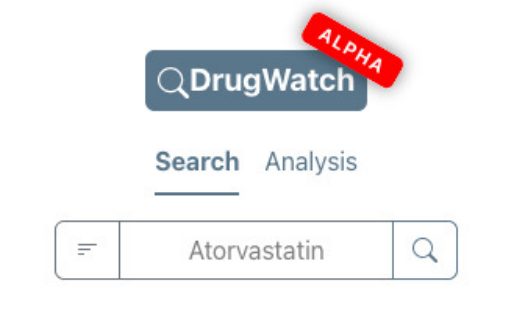}
    \caption{Screenshot of the main page (search box).}
    \label{fig:search_page}
\end{figure}

Figure \ref{fig:search_page} presents the visual of the main search page. When users type in the names of drugs or side effects, a dropdown window will automatically pop up with suggested search items. Users add search terms by clicking on the suggested items. They can also type and select multiple drug or side effect names in succession for a combined search.

\begin{figure}[!h]
    \centering
    \includegraphics[width=0.9\linewidth]{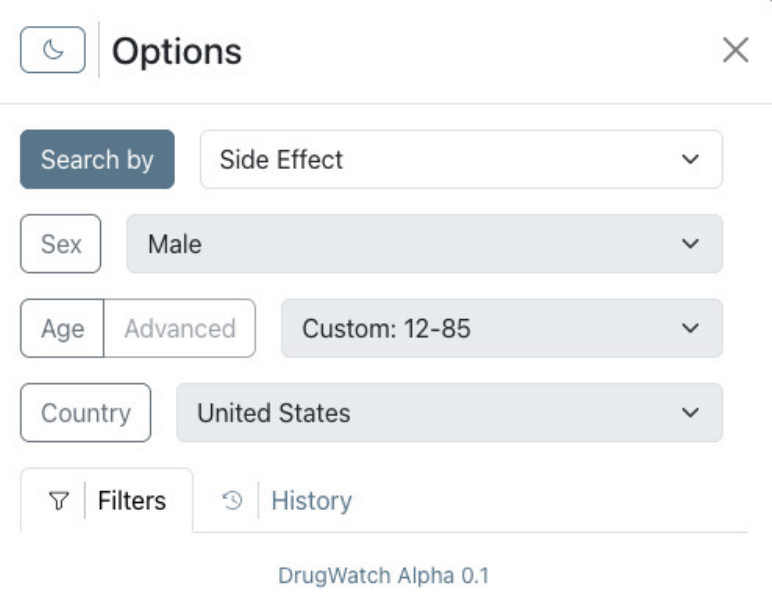}
    \caption{Screenshot of the search options.}
    \label{fig:search_options}
\end{figure}

Figure \ref{fig:search_options} demonstrates the search criteria that can be specified by users. First, users must set whether to perform the search based on `Generic Name', `Brand Name' (for drugs), or `Side Effect'. Subsequently, users can optionally add filtering conditions, including the patient's age, gender, and nationality mentioned in the report. Furthermore, we allow users to retain their last five search histories, as well as choose between a bright or dark UI style according to their preference.

\begin{figure}[!h]
    \centering
    \includegraphics[width=1\linewidth]{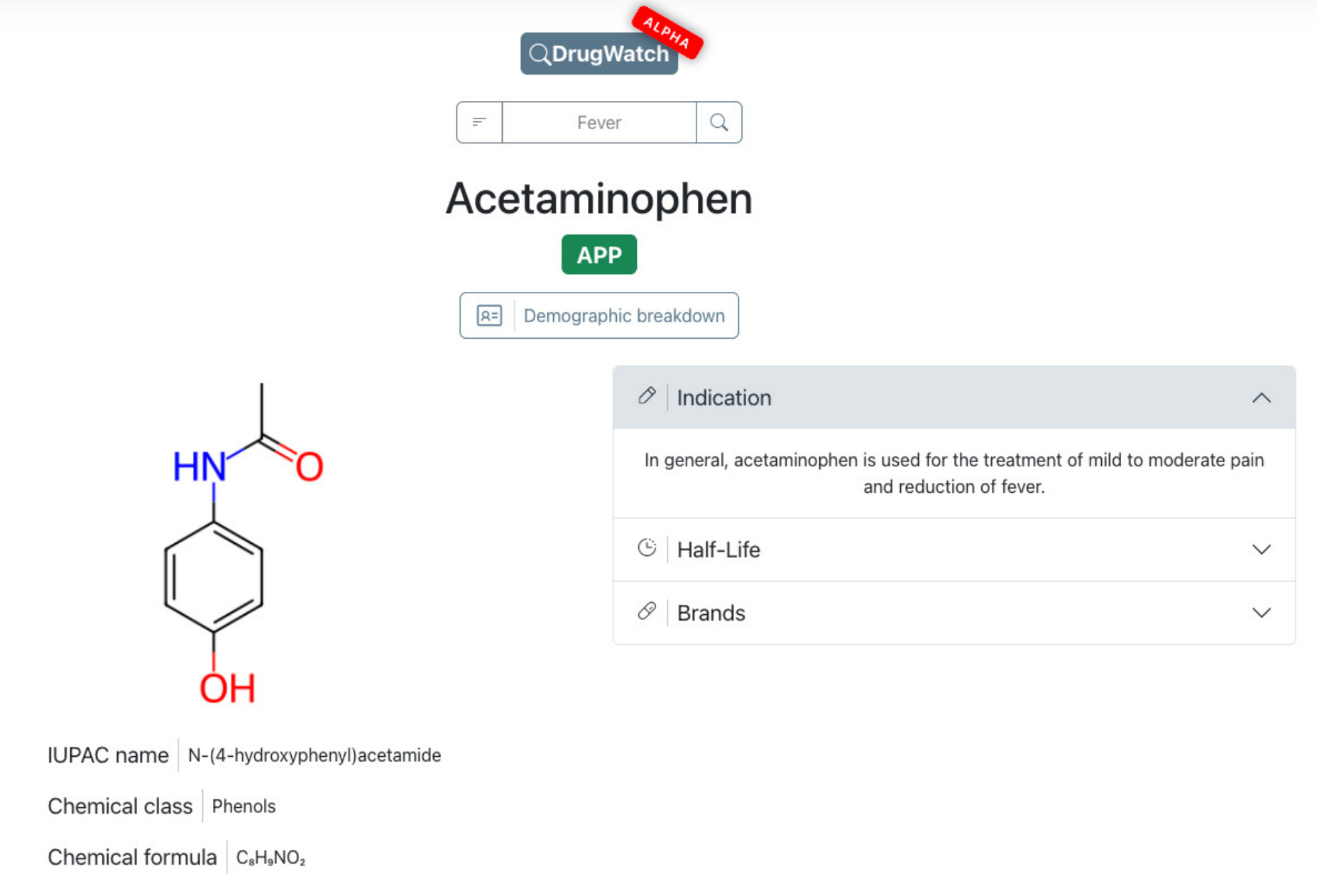}
    \caption{Screenshot of drug information display.}
    \label{fig:drug_info}
\end{figure}

Figure \ref{fig:drug_info} shows the essential drug information presented to the users. In addition to fundamental chemical and pharmaceutical details, our system incorporates tags to specify a drug's status, including approval (APP), investigation (INV), illegality (ILL), historical veterinary approval (VET), withdrawal (WIT), nutraceutical designation (NUT), experimental status (EXP), or reported side effects (SID).

\begin{figure}[!h]
    \centering
    \includegraphics[width=1\linewidth]{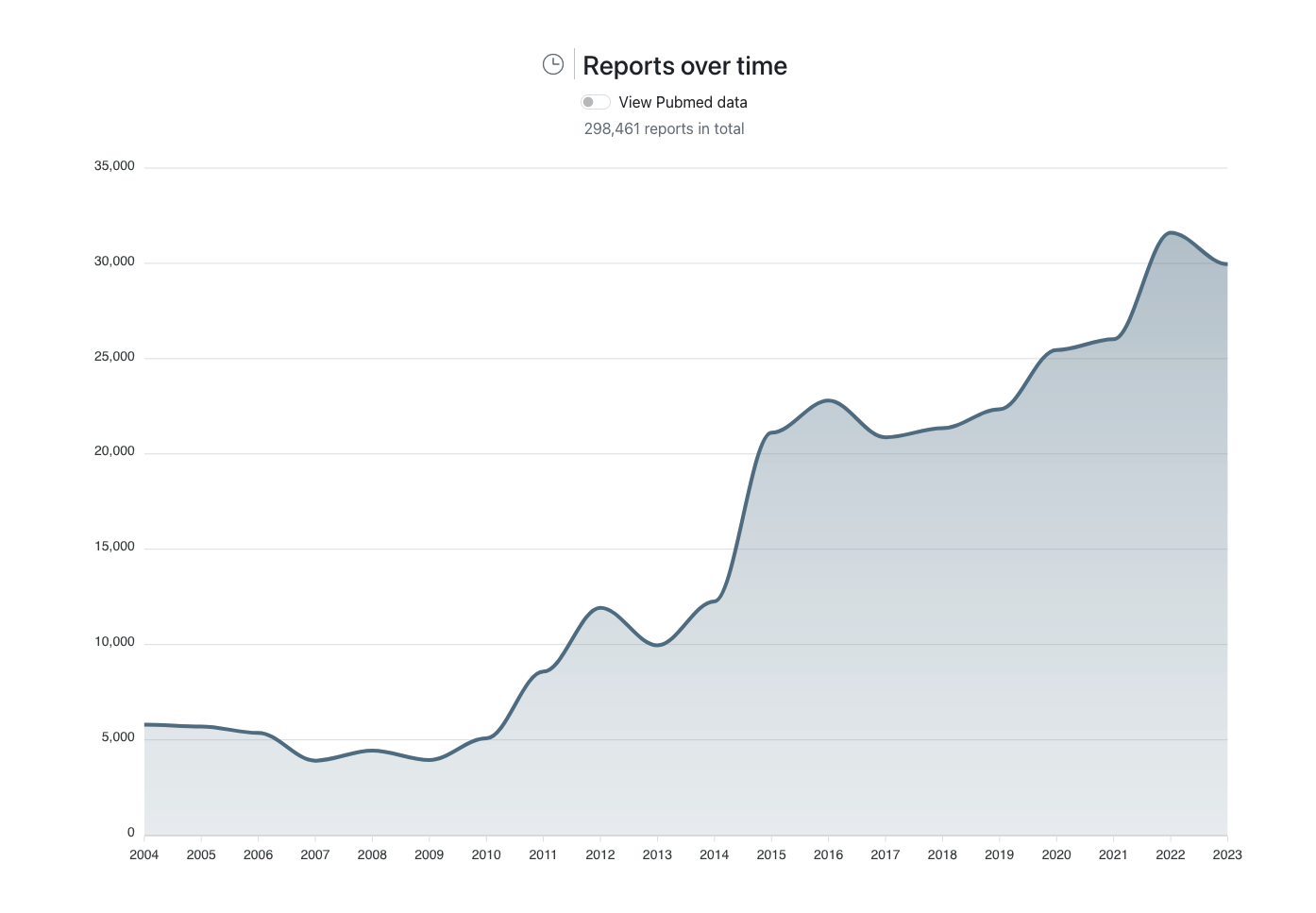}
    \caption{The line chart shows the counts of reports over time for Acetaminophen.}
    \label{fig:time_series}
\end{figure}

Figure \ref{fig:time_series} shows a graph showing changes in the number of reports over the years. If the user specifies a demographic filter when searching, the number of reports in the curve is after filtered. Users can also easily click the adjacent button to view the curves without applying any demographic filter for comparison. The chart is interactive and will render varied report counts depending on the mouse position on the chart.

Users have various ways to view demographic information for a search term. Firstly, if a user is searching for a drug (or side effect), they can access the demographic information page for that drug (or side effect) by clicking on a button at the top of the results page. Additionally, if a user is interested in demographic information for the most associated side effects (or drugs) when searching for a drug (or side effect), they can directly click on the bar chart or word cloud to reach the demographic information page for that side effect (or drug). When the user exits the pop-up demographic information window, the current search results remain visible.

\begin{figure}[!h]
    \centering
    \includegraphics[width=1\linewidth]{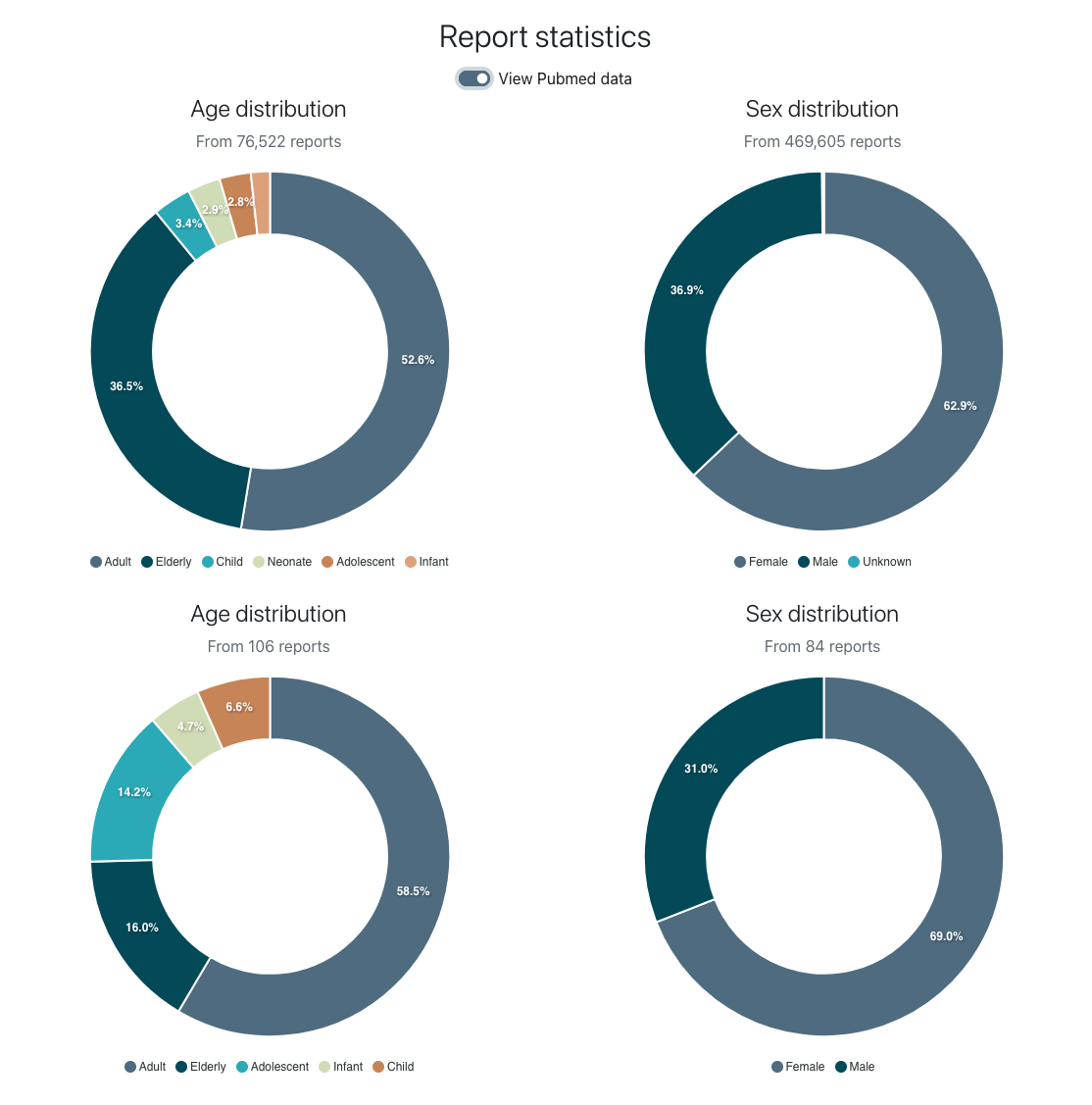}
    \caption{Overall demographic distribution of reports related to Acetaminophen. (Above for FAERS data, and bottom for PubMed data.)}
    \label{fig:demographic_distribution}
\end{figure}

Figure \ref{fig:demographic_distribution} illustrates the overall distribution of age and gender groups associated with reports related to the searched drug (or side effect). We will label the total number of relevant reports above the pie chart and annotate the proportions of each group on the chart. When hovering over the pie chart, users can also view the number of reports for each subgroup. Users could also view the graph for PubMed data by turning on the button. 

\begin{figure}[!h]
    \centering
    \includegraphics[width=1\linewidth]{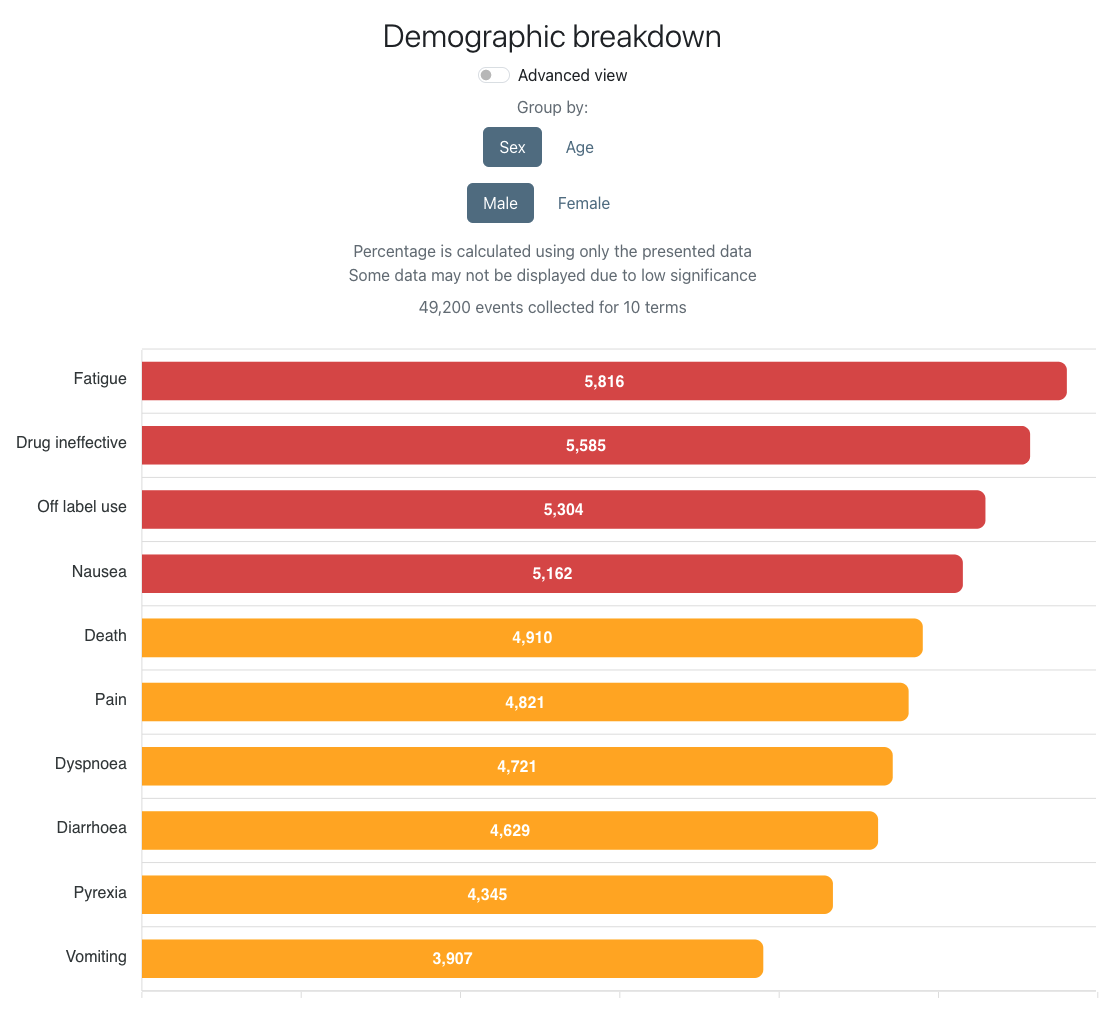}
    \caption{Basic view of demographic information breakdown for Acetaminophen.}
    \label{fig:breakdown_basic}
\end{figure}

Figure \ref{fig:breakdown_basic} shows the basic view of demographic information breakdown. Users can choose to view the top 10 side effects (or drugs) related to the drug (or side effect) for any gender group or age group. The bar chart will display their report counts. When the user hovers the mouse over a bar, the user can view the proportion of reports related to this side effect among the top ten side effects. Limited by the calling method of the OpenFDA API, we are currently unable to display the proportion of this side effect among all side effects.

\begin{figure}[!h]
    \centering
    \includegraphics[width=1\linewidth]{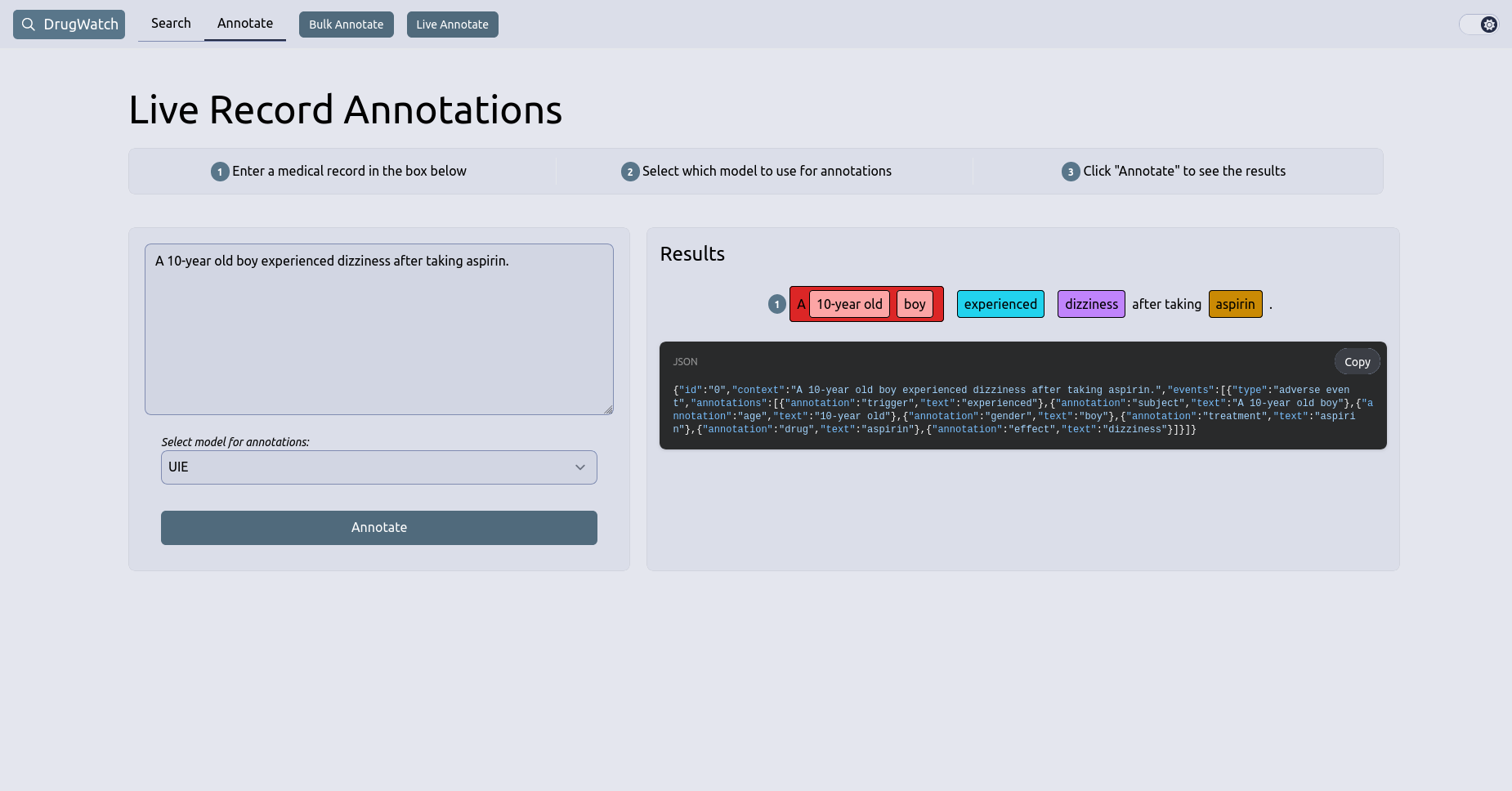}
    \caption{Screenshot of \textit{DrugWatch Search} live annotation page.}
    \label{fig:live_annotate}
\end{figure}

Figure \ref{fig:live_annotate} presents the interface where the user can input a single sentence and instantly check the model annotation results. 

\begin{figure}[!h]
    \centering
    \includegraphics[width=1\linewidth]{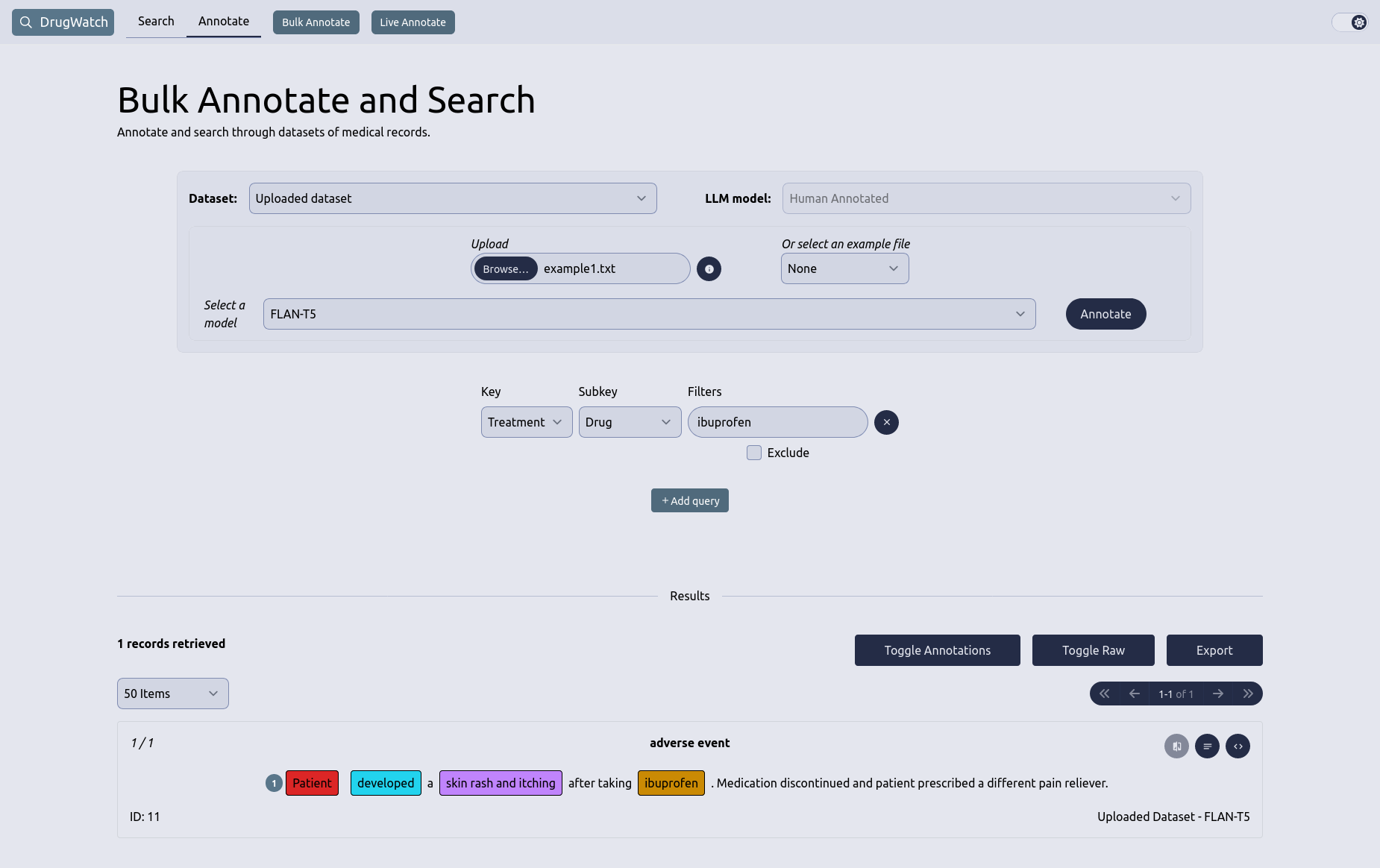}
    \caption{Screenshot of \textit{DrugWatch Search} bulk annotation page.}
    \label{fig:bulk_annotate}
\end{figure}

Figure \ref{fig:bulk_annotate} shows the bulk annotate interface where users can batch upload and annotate their data. In this interface, users can choose to view the built-in data set or upload their own data, select the model they want to use, and freely add conditions for filtering results. We distinguish different argument types with different colours to provide a more intuitive visual effect.

\end{document}